\documentclass{PoS}

\usepackage{amssymb}
\usepackage{amsthm}
\usepackage{amsmath}
\usepackage{booktabs}
\usepackage{bbm}
\DeclareMathOperator{\tr}{tr}

\numberwithin{equation}{section}
\usepackage{url}

\title{Background field method in the gradient flow}

\ShortTitle{Background field method in the gradient flow}

\author{\speaker{Hiroshi Suzuki}\thanks{The work of H.~S. is supported in part
by Grant-in-Aid for Scientific Research~23540330.}\\
Department of Physics, Kyushu University,
744 Motooka, Nishi-ku, Fukuoka, 819-0395, Japan\thanks{Address after October
1st.}
        E-mail: \email{hsuzuki@phys.kyushu-u.ac.jp}}

\abstract{%
The Yang--Mills gradient flow and its extension to the fermion field provide a
very general method to obtain renormalized observables in gauge theory. The
method is applicable also with non-perturbative regularization such as lattice.
The gradient flow thus offers useful probes to study non-perturbative dynamics
of gauge theory. In this work, aiming at possible simplification in
perturbative calculations associated with the gradient flow, a modification of
the gauge-fixed version of the flow equation, which preserves
gauge covariance under the background gauge transformation, is proposed. This
formulation allows for example a very quick one-loop calculation of the small
flow time expansion of a composite operator that is relevant to the
construction of a lattice energy--momentum tensor. Some details of the
calculation, which have not been given elsewhere, are presented.
}

\FullConference{The 33rd International Symposium on Lattice Field Theory\\
		14 -18 July 2015\\
		Kobe International Conference Center, Kobe, Japan*}

\begin{document}

\section{Introduction}
\label{sec:1}
The Yang--Mills gradient flow~\cite{Luscher:2010iy,Luscher:2011bx} and its
extension to the fermion field~\cite{Luscher:2013cpa} provide a very general
method to obtain renormalized observables in gauge theory. The method is
applicable also with non-perturbative regularization such as lattice. The
gradient flow thus offers useful probes to study non-perturbative dynamics of
gauge theory. See Ref.~\cite{Ramos:2015dla} for a recent review.

As noted in~Ref.~\cite{Luscher:2010iy}, for perturbative consideration of the
gradient flow, it is useful to introduce a ``gauge fixing term'' to the
flow equation that breaks gauge covariance. This gauge fixing term gives rise
to a Gaussian damping factor also for gauge degrees of freedom which ensures a
good convergence property of momentum integrals; this then facilitates
perturbative consideration of the gradient flow such as the proof of its
renormalizability~\cite{Luscher:2011bx,Luscher:2013cpa}.

In this work, aiming at possible simplification in perturbative calculations
associated with the gradient flow, a modification of (the gauge-fixed version
of) the flow equation of~Ref.~\cite{Luscher:2010iy} is proposed. Our proposal
is
\begin{equation}
   \partial_tB_\mu(t,x)=D_\nu G_{\nu\mu}(t,x)
   +\alpha_0D_\mu\Hat{D}_\nu b_\nu(t,x),\qquad
   B_\mu(t=0,x)=A_\mu(x),
\label{eq:(1.1)}
\end{equation}
where $t>0$ is the flow time which parametrizes the flow of the gauge
field~$B_\mu(t,x)$; $D_\nu$ and~$G_{\nu\mu}$ stand for the covariant derivative
and field strength of the flowed gauge field, respectively. The term being
proportional to~$\alpha_0$ is the ``gauge fixing term'' mentioned above. To
define this term, we decompose the gauge fields into the \emph{background\/}
and~\emph{quantum\/} parts as $A_\mu(x)=\Hat{A}_\mu(x)+a_\mu(x)$ and,
correspondingly, $B_\mu(t,x)=\Hat{B}_\mu(t,x)+b_\mu(t,x)$. Then
in~Eq.~\eqref{eq:(1.1)},
\begin{equation}
   \Hat{D}_\mu=\partial_\mu+[\Hat{B}_\mu,\cdot]
\label{eq:(1.2)}
\end{equation}
is the covariant derivatives with respect to the background
field~$\Hat{B}_\mu(t,x)$. The idea is that, as the conventional background
field method~\cite{'tHooft:1975vy,DeWitt:1980jv,Boulware:1980av,Abbott:1980hw,%
Ichinose:1981uw}, the ``gauge fixing term'' is designed so that covariance
under the background gauge transformation is preserved; the quantum fields
transform as the adjoint representation under the background gauge
transformation. In addition to~Eq.~\eqref{eq:(1.1)}, we postulate that the
background field obeys its own flow equation:
\begin{equation}
   \partial_t\Hat{B}_\mu(t,x)=\Hat{D}_\nu\Hat{G}_{\nu\mu}(t,x),\qquad
   \Hat{B}_\mu(t=0,x)=\Hat{A}_\mu(x),
\label{eq:(1.3)}
\end{equation}
where $\Hat{G}_{\nu\mu}(t,x)$ is the field strength of the background
field~$\Hat{B}_\mu(t,x)$.

Since the present study is already published in~Ref.~\cite{Suzuki:2015bqa}, in
these proceedings, I will reproduce some materials which were not explicitly
given in~Ref.~\cite{Suzuki:2015bqa}. In particular, since the most interesting
result obtained in~Ref.~\cite{Suzuki:2015bqa} is a one-loop calculation of the
small flow time expansion of the composite
operator~$G_{\mu\rho}^a(t,x)G_{\nu\rho}^a(t,x)$, which is relevant to the
construction of a lattice energy--momentum tensor via the gradient
flow~\cite{Suzuki:2013gza,Makino:2014taa} (see also
Ref.~\cite{Asakawa:2013laa}), we will present some details of the
calculation\footnote{Our calculational scheme has been originally inspired
by the calculation in~Ref.~\cite{Fujikawa:1993xv}.} which were omitted
in~Ref.~\cite{Suzuki:2015bqa}; we do not treat the fermion flow in the present
article. In this way, I hope that the present article becomes complementary
to~Ref.~\cite{Suzuki:2015bqa}. Our notational convention is identical to that
of~Ref.~\cite{Suzuki:2015bqa}; in particular, generators of the gauge group are
normalized as~$\tr(T^aT^b)=(-1/2)\delta^{ab}$.

\section{Small flow time expansion relevant to the energy--momentum tensor}
\label{sec:2}
The initial condition in the flow equation~\eqref{eq:(1.1)}, $A_\mu(x)$, is
subject of the functional integral with the Boltzmann weight, specified by the
Yang--Mills and background-gauge-fixing actions:
\begin{equation}
   S+S_{\mathrm{gf}}
   =-\frac{1}{2g_0^2}\int d^Dx\,\tr\left[F_{\mu\nu}(x)F_{\mu\nu}(x)\right]
   -\frac{\lambda_0}{g_0^2}\int d^Dx\,
   \tr\left[\Hat{D}_\mu a_\mu(x)\Hat{D}_\nu a_\nu(x)\right],
\label{eq:(2.1)}
\end{equation}
here $D\equiv4-2\epsilon$ is the spacetime dimension. The Faddeev--Popov
ghost action corresponding to this background gauge fixing is given by
\begin{equation}
   S_{c\Bar{c}}=\frac{2}{g_0^2}\int d^Dx\,
   \tr\left[\Bar{c}(x)\Hat{D}_\mu D_\mu c(x)\right].
\label{eq:(2.2)}
\end{equation}
From these, the tree-level propagators \emph{in the presence of the
background field} are given by, for the Feynman gauge~$\lambda_0=1$,
\begin{equation}
   \left\langle a_\mu^a(x)a_\nu^b(y)\right\rangle_0
    =-g_0^2\left(
   \frac{1}{\Hat{\Delta}_x}
   \right)_{\mu\nu}^{ab}\delta(x-y),\qquad
   \left\langle c^a(x)\Bar{c}^b(y)\right\rangle_0
   =-g_0^2\left(\frac{1}{\Hat{\mathcal{D}}_x^2}\right)^{ab}\delta(x-y),
\label{eq:(2.3)}
\end{equation}
where
\begin{equation}
   \Hat{\mathcal{D}}_\mu^{ab}
   \equiv\delta^{ab}\partial_\mu+\Hat{A}_\mu^c(x)f^{acb}
\label{eq:(2.4)}
\end{equation}
is the background covariant derivative in the adjoint representation;
$(\Hat{\mathcal{D}}^2)^{ab}%
=\Hat{\mathcal{D}}_\mu^{ac}\Hat{\mathcal{D}}_\mu^{cb}$ and
\begin{equation}
   \Hat{\Delta}_{\mu\nu}^{ab}
   =(\Hat{\mathcal{D}}^2)^{ab}\delta_{\mu\nu}
   +2\Hat{\mathcal{F}}_{\mu\nu}^{ab},\qquad
   \Hat{\mathcal{F}}_{\mu\nu}^{ab}(x)
   \equiv\Hat{F}_{\mu\nu}^c(x)f^{acb}.
\label{eq:(2.5)}
\end{equation}

In what follows, as~Ref.~\cite{Suzuki:2015bqa}, we assume that the background
field~$\Hat{A}_\mu(x)$ obeys the Yang--Mills equation of motion:
$\Hat{D}_\nu\Hat{F}_{\nu\mu}(x)=0$. Eq.~\eqref{eq:(1.3)} then implies that the
background gauge field \emph{does not flow}, $\Hat{B}(t,x)=\Hat{A}(x)$. This
assumption considerably simplifies all the expressions and, in particular, the
tree-level propagator of the flowed quantum field in the presence of the
background field, for the ``Feynman gauge'' $\alpha_0=1$,\footnote{It can be
shown that any gauge invariant quantity that does not contain the flow time
derivative is independent of~$\alpha_0$~\cite{Suzuki:2015bqa}.} is given
by~\cite{Suzuki:2015bqa}
\begin{equation}
   \left\langle b_\mu^a(t,x)b_\nu^b(s,y)\right\rangle_0
   =-g_0^2\left[
   e^{(t+s)\Hat{\Delta}_x}\,
   \frac{1}{\Hat{\Delta}_x}
   \right]_{\mu\nu}^{ab}\delta(x-y).
\label{eq:(2.6)}
\end{equation}

Now, for the construction of a lattice energy--momentum tensor
in~Refs.~\cite{Suzuki:2013gza,Makino:2014taa}, one has to find the coefficients
$\zeta_{11}(t)$ and~$\zeta_{12}(t)$ in the small flow time
expansion~\cite{Luscher:2011bx} of the form,
\begin{align}
   &G_{\mu\rho}^a(t,x)G_{\nu\rho}^a(t,x)
\notag\\
   &\stackrel{t\to0}{\sim}
   \left\langle G_{\mu\rho}^a(t,x)G_{\nu\rho}^a(t,x)\right\rangle
   +\zeta_{11}(t)F_{\mu\rho}^a(x)F_{\nu\rho}^a(x)
   +\zeta_{12}(t)\delta_{\mu\nu}F_{\rho\sigma}^a(x)F_{\rho\sigma}^a(x)
   +O(t).
\label{eq:(2.7)}
\end{align}

For our background-quantum decomposition,
\begin{align}
   G_{\mu\rho}^aG_{\nu\rho}^a
   &=\Hat{F}_{\mu\rho}^a\Hat{F}_{\nu\rho}^a
   +\Hat{F}_{\mu\rho}^a\left(\Hat{D}_\nu b_\rho-\Hat{D}_\rho b_\nu\right)^a
   +\left(\Hat{D}_\mu b_\rho-\Hat{D}_\rho b_\mu\right)^a\Hat{F}_{\nu\rho}^a
\notag\\
   &\qquad{}
   +\left(\Hat{D}_\mu b_\rho-\Hat{D}_\rho b_\mu\right)^a
   \left(\Hat{D}_\nu b_\rho-\Hat{D}_\rho b_\nu\right)^a
   +\Hat{F}_{\mu\rho}^a[b_\nu,b_\rho]^a
   +[b_\mu,b_\rho]^a\Hat{F}_{\nu\rho}^a+O(b^3),
\label{eq:(2.8)}
\end{align}
where we have used $\Hat{B}(t,x)=\Hat{A}(x)$. A similar expansion holds also
for~$F_{\mu\rho}^aF_{\nu\rho}^a$. Thus, in the tree-level, in which the quantum
fields are treated as zero, we have
\begin{equation}
   \left\langle
   G_{\mu\rho}^a(t,x)G_{\nu\rho}^a(t,x)
   \right\rangle
   =\Hat{F}_{\mu\rho}^a(x)\Hat{F}_{\nu\rho}^a(x),\qquad
   \left\langle
   F_{\mu\rho}^a(x)F_{\nu\rho}^a(x)
   \right\rangle
   =\Hat{F}_{\mu\rho}^a(x)\Hat{F}_{\nu\rho}^a(x),
\label{eq:(2.9)}
\end{equation}
and from these, $G_{\mu\rho}^a(t,x)G_{\nu\rho}^a(t,x)\stackrel{t\to0}{\sim}%
F_{\mu\rho}^a(x)F_{\nu\rho}^a(x)+O(t)$ in the tree level.

For the one-loop calculation, as noted in~Refs.~\cite{Makino:2014taaa}, it is
convenient to consider the correlation function
\begin{equation}
   \left\langle
   G_{\mu\rho}^a(t,x)G_{\nu\rho}^a(t,x)
   -F_{\mu\rho}^a(x)F_{\nu\rho}^a(x)\right\rangle,
\label{eq:(2.10)}
\end{equation}
because possible infrared divergences are cancelled out in this difference.
Noting Eq.~\eqref{eq:(2.8)}, we first consider the tadpole contribution
to~Eq.~\eqref{eq:(2.10)}
\begin{align}
   \Hat{F}_{\mu\rho}^a
   \left(\Hat{D}_\nu\left\langle b_\rho-a_\rho\right\rangle
   -\Hat{D}_\rho\left\langle b_\nu-a_\nu\right\rangle\right)^a
   +\left(\Hat{D}_\mu\left\langle b_\rho-a_\rho\right\rangle
   -\Hat{D}_\rho\left\langle b_\mu-a_\mu\right\rangle\right)^a
   \Hat{F}_{\nu\rho}^a.
\label{eq:(2.11)}
\end{align}
For the one-point function $\langle b_\mu^a(t,x)-a_\mu^a(x)\rangle$ in the
presence of the background field, there are two types of one-loop diagrams.
The first type is tadpole diagrams containing the vertex at the vanishing flow
time. The contribution of this type is
\begin{align}
   \left\langle b_\mu^a(t,x)-a_\mu^a(x)\right\rangle
   &=g_0^2\left[\left(e^{t\Hat{\Delta}_x}-1\right)
   \frac{1}{\Hat{\Delta}_x}\right]_{\mu\nu}^{ab}f^{bcd}
\notag\\
   &\qquad{}
   \times\lim_{y\to x}\left[
   -2\left(\frac{1}{\Hat{\Delta}_x}\right)_{\rho\nu}^{ce}
   \Hat{\mathcal{D}}_{x\rho}^{ed}
   +\left(\frac{1}{\Hat{\Delta}_x}\right)_{\rho\rho}^{ce}
   \Hat{\mathcal{D}}_{x\nu}^{ed}
   +\left(\frac{1}{\Hat{\Delta}_x}
   -\frac{1}{\Hat{\mathcal{D}}_x^2}\right)_{\nu\rho}^{ce}
   \Hat{\mathcal{D}}_{x\rho}^{ed}\right]\delta(x-y),
\label{eq:(2.12)}
\end{align}
where the last term is the contribution of the ghost loop. In this expression,
the flow time~$t$ is contained only in the first factor and this is
obviously~$O(t)$. Thus, this first type does not contribute to the $O(t^0)$
terms in~Eq.~\eqref{eq:(2.7)}.

The second type of diagram is the contribution of the vertex at non-zero
flow time (flow vertex):
\begin{equation}
   \left\langle b_\mu^a(t,x)\right\rangle
   =g_0^2\int_0^tds\,\left[e^{(t-s)\Hat{\Delta}_x}\right]_{\mu\nu}^{ab}f^{bcd}
   \lim_{y\to x}\left[
   2\left(e^{2s\Hat{\Delta}_x}\frac{1}{\Hat{\Delta}_x}\right)_{\rho\nu}^{ce}
   \Hat{\mathcal{D}}_{x\rho}^{ed}
   -\left(e^{2s\Hat{\Delta}_x}\frac{1}{\Hat{\Delta}_x}\right)_{\rho\rho}^{ce}
   \Hat{\mathcal{D}}_{x\nu}^{ed}
   \right]\delta(x-y).
\label{eq:(2.13)}
\end{equation}
To study the $t\to0$ behavior of this, we set
$\delta(x-y)=\int\frac{d^Dp}{(2\pi)^D}\,e^{ipx}e^{-ipy}$, and moves the plain
wave~$e^{ipx}$ to the most left-hand side under the limit~$\lim_{y\to x}$ as the
Fujikawa method~\cite{Fujikawa:1979ay,Fujikawa:1980eg}. For this, we note
\begin{equation}
   \Hat{\mathcal{D}}_\mu e^{ipx}
   =e^{ipx}(ip_\mu+\Hat{\mathcal{D}}_\mu),
\label{eq:(2.14)}
\end{equation}
and rescale the integration variable as~$p_\mu\to p_\mu/\sqrt{s}$. Then,
Eq.~\eqref{eq:(2.13)} becomes
\begin{align}
   \left\langle b_\mu^a(t,x)\right\rangle
   &=g_0^2\int_0^tds\,\left[e^{(t-s)\Hat{\Delta}_x}\right]_{\mu\nu}^{ab}f^{bcd}
   s^{-D/2+1/2}\int\frac{d^Dp}{(2\pi)^D}\,e^{-2p^2}
\notag\\
   &\qquad{}
   \times
   \Biggl[
   2\left(e^{4i\sqrt{s}p\cdot\Hat{\mathcal{D}}+2s\Hat{\Delta}}
   \frac{1}{-p^2+2i\sqrt{s}p\cdot\Hat{\mathcal{D}}+s\Hat{\Delta}}
   \right)_{\rho\nu}^{ce}
   \left(ip_\rho+\Hat{\mathcal{D}}_\rho\right)^{ed}
\notag\\
   &\qquad\qquad{}
   -\left(e^{4i\sqrt{s}p\cdot\Hat{\mathcal{D}}+2s\Hat{\Delta}}
   \frac{1}{-p^2+2i\sqrt{s}p\cdot\Hat{\mathcal{D}}+s\Hat{\Delta}}
   \right)_{\rho\rho}^{ce}
   \left(ip_\nu+\Hat{\mathcal{D}}_\nu\right)^{ed}
   \Biggr].
\label{eq:(2.15)}
\end{align}
A straightforward expansion with respect to~$\sqrt{s}$ then yields
\begin{equation}
   \left\langle b_\mu^a(t,x)\right\rangle
   =g_0^2f^{abc}\Hat{\mathcal{D}}_\mu^{bc}
   (2-D)\int_0^tds\,s^{-D/2+1}\int\frac{d^Dp}{(2\pi)^D}\,e^{-2p^2}
   \left[\frac{4}{D}+\frac{1}{p^2}\left(-1+\frac{2}{D}\right)\right]
   +O(t^{1+\epsilon}),
\label{eq:(2.16)}
\end{equation}
and this momentum integration identically vanishes for any~$D$. This shows that
one-loop tadpole diagrams do not contribute to the $O(t^0)$ terms
in~Eq.~\eqref{eq:(2.7)}.

Next, we consider one-loop diagrams which arise from the contraction of quantum
fields in the last line of~Eq.~\eqref{eq:(2.8)} by
propagators~\eqref{eq:(2.3)} and~\eqref{eq:(2.6)}. A procedure being similar to
that led to~Eq.~\eqref{eq:(2.15)} yields Eq.~(3.16)
of~Ref.~\cite{Suzuki:2015bqa}; it is
\begin{align}
   &\left\langle
   \left.G_{\mu\rho}^a(t,x)G_{\nu\rho}^a(t,x)\right|_{O(b^2)}
   -\left.F_{\mu\rho}^a(x)F_{\nu\rho}^a(x)\right|_{O(a^2)}
   \right\rangle
\notag\\
   &=2g_0^2\int_0^td\xi\,\xi^{-D/2}
   \int\frac{d^Dp}{(2\pi)^D}\,e^{-2p^2}
\notag\\
   &\qquad{}\times
   \tr\biggl[
   \xi^{-1}\mathcal{P}_{\mu\alpha,\nu\delta,\beta\gamma}
   \left(ip_\alpha+\sqrt{\xi}\Hat{\mathcal{D}}_\alpha\right)
   \left(e^{4i\sqrt{\xi}p\cdot\Hat{\mathcal{D}}+2\xi\Hat{\Delta}}
   \right)_{\beta\gamma}
   \left(ip_\delta+\sqrt{\xi}\Hat{\mathcal{D}}_\delta\right)
\notag\\
   &\qquad\qquad\qquad\qquad{}
   +\Hat{\mathcal{F}}_{\mu\rho}(x)
   \left(e^{4i\sqrt{\xi}p\cdot\Hat{\mathcal{D}}+2\xi\Hat{\Delta}}
   \right)_{\rho\nu}
   +\Hat{\mathcal{F}}_{\nu\rho}(x)
   \left(e^{4i\sqrt{\xi}p\cdot\Hat{\mathcal{D}}+2\xi\Hat{\Delta}}
   \right)_{\rho\mu}
   \biggr],
\label{eq:(2.17)}
\end{align}
where
\begin{equation}
   \mathcal{P}_{\mu\alpha,\nu\delta,\beta\gamma}
   \equiv\delta_{\mu\alpha}\delta_{\nu\delta}\delta_{\beta\gamma}
   -\delta_{\mu\alpha}\delta_{\nu\gamma}\delta_{\beta\delta}
   -\delta_{\mu\beta}\delta_{\nu\delta}\delta_{\alpha\gamma}
   +\delta_{\mu\beta}\delta_{\nu\gamma}\delta_{\alpha\delta}.
\label{eq:(2.18)}
\end{equation}
We note that, in the present background-gauge-covariant formulation, this
compact expression~\eqref{eq:(2.17)} contains all the information equivalent to
the tedious diagrammatic expansion computed in~Ref.~\cite{Suzuki:2013gza}.

We first compute the second term of~Eq.~\eqref{eq:(2.17)}:
\begin{equation}
   2g_0^2\int_0^td\xi\,\xi^{-D/2}
   \int\frac{d^Dp}{(2\pi)^D}\,e^{-2p^2}
   \tr\left[
   \Hat{\mathcal{F}}_{\mu\rho}(x)
   \left(e^{4i\sqrt{\xi}p\cdot\Hat{\mathcal{D}}+2\xi\Hat{\Delta}}
   \right)_{\rho\nu}\right].
\label{eq:(2.19)}
\end{equation}
The expansion of this expression for~$t\to0$ is easy because, as noted
in~Ref.~\cite{Suzuki:2015bqa}, only terms symmetric
under~$\mu\leftrightarrow\nu$ contribute by definition.
From~Eqs.~(A1) and~(A2) of~Ref.~\cite{Suzuki:2015bqa}, we immediately see that
the expansion yields
\begin{align}
   &8g_0^2\int_0^td\xi\,\xi^{-D/2+1}
   \int\frac{d^Dp}{(2\pi)^D}\,e^{-2p^2}
   \tr\left[
   \Hat{\mathcal{F}}_{\mu\rho}(x)\Hat{\mathcal{F}}_{\rho\nu}(x)\right]
   +O(t^{3-D/2})
\notag\\
   &=\frac{g_0^2}{(4\pi)^2}\frac{(8\pi t)^\epsilon}{\epsilon}
   2\tr\left[\Hat{\mathcal{F}}(x)^2\right]_{\mu\nu}
   +O(t^{1+\epsilon}).
\label{eq:(2.20)}
\end{align}
The last term in~Eq.~\eqref{eq:(2.17)} gives rise to the same contribution.

The computation of the first term of~Eq.~\eqref{eq:(2.17)} is somewhat
complicated. We first consider the expression without the
factor~$\mathcal{P}_{\mu\alpha,\nu\delta,\beta\gamma}$;
\begin{equation}
   2g_0^2\int_0^td\xi\,\xi^{-D/2-1}
   \int\frac{d^Dp}{(2\pi)^D}\,e^{-2p^2}
   \tr\left[
   \left(ip_\alpha+\sqrt{\xi}\Hat{\mathcal{D}}_\alpha\right)
   \left(e^{4i\sqrt{\xi}p\cdot\Hat{\mathcal{D}}+2\xi\Hat{\Delta}}
   \right)_{\beta\gamma}
   \left(ip_\delta+\sqrt{\xi}\Hat{\mathcal{D}}_\delta\right)\right].
\label{eq:(2.21)}
\end{equation}
We then use Eqs.~(A1)--(A3) of~Ref.~\cite{Suzuki:2015bqa}. We can neglect
$O(\xi^{-D/2})$~terms
in the integrand, because these terms give rise to $O(t^{-1})$ terms
in~Eq.~\eqref{eq:(2.17)} for~$D\to4$ which must be absent from gauge invariance.
After the expansion, we carry out the momentum integration. Then since
\begin{equation}
   \mathcal{P}_{\mu\alpha,\nu\delta,\beta\gamma}X_{\alpha\delta}Y_{\beta\gamma}
   =X_{\mu\nu}Y_{\rho\rho}-X_{\mu\rho}Y_{\rho\nu}-X_{\rho\nu}Y_{\mu\rho}
   +X_{\rho\rho}Y_{\mu\nu},
\label{eq:(2.22)}
\end{equation}
for the combination~\eqref{eq:(2.18)}, if either~$X$ or~$Y$ is symmetric under
the exchange of indices and the other is anti-symmetric, the
product~$\mathcal{P}_{\mu\alpha,\nu\delta,\beta\gamma}X_{\alpha\delta}Y_{\beta\gamma}$
is anti-symmetric under~$\mu\leftrightarrow\nu$. Such term should not
contribute to~Eq.~\eqref{eq:(2.17)}. Thus, what we have to retain are only
combinations such that both $X$ and~$Y$ are symmetric or anti-symmetric under
the exchange of indices. This observation greatly simplifies our task. After
some calculation, we have for~Eq.~\eqref{eq:(2.21)},
\begin{align}
   &\frac{g_0^2}{(4\pi)^2}\frac{(8\pi t)^\epsilon}{1-\epsilon/2}
   \frac{1}{16t^2}\delta_{\alpha\delta}\delta_{\beta\gamma}\dim G
\notag\\
   &\qquad{}
   +\frac{g_0^2}{(4\pi)^2}\frac{(8\pi t)^\epsilon}{\epsilon}
   \tr\biggl\{
   -\delta_{\alpha\delta}\Hat{\mathcal{F}}(x)^2_{\beta\gamma}
   -[\Hat{\mathcal{D}}_\alpha,\Hat{\mathcal{D}}_\delta]
   \Hat{\mathcal{F}}(x)_{\beta\gamma}
\notag\\
   &\qquad\qquad{}
   +\frac{1}{12}\delta_{\beta\gamma}\Bigl[
   \Hat{\mathcal{D}}_\alpha\Hat{\mathcal{D}}_\varepsilon\Hat{\mathcal{D}}_\delta
   \Hat{\mathcal{D}}_\varepsilon
   +\Hat{\mathcal{D}}_\delta\Hat{\mathcal{D}}_\varepsilon\Hat{\mathcal{D}}_\alpha
   \Hat{\mathcal{D}}_\varepsilon
   -\Hat{\mathcal{D}}_\alpha\Hat{\mathcal{D}}^2\Hat{\mathcal{D}}_\delta
   -\Hat{\mathcal{D}}_\delta\Hat{\mathcal{D}}^2\Hat{\mathcal{D}}_\alpha
\notag\\
   &\qquad\qquad\qquad\qquad\qquad{}
   -\Hat{\mathcal{D}}_\varepsilon\Hat{\mathcal{D}}_\alpha\Hat{\mathcal{D}}_\delta
   \Hat{\mathcal{D}}_\varepsilon
   -\Hat{\mathcal{D}}_\varepsilon\Hat{\mathcal{D}}_\delta\Hat{\mathcal{D}}_\alpha
   \Hat{\mathcal{D}}_\varepsilon
   +\Hat{\mathcal{D}}_\varepsilon\Hat{\mathcal{D}}_\alpha
   \Hat{\mathcal{D}}_\varepsilon\Hat{\mathcal{D}}_\delta
   +\Hat{\mathcal{D}}_\varepsilon\Hat{\mathcal{D}}_\delta
   \Hat{\mathcal{D}}_\varepsilon\Hat{\mathcal{D}}_\alpha
\notag\\
   &\qquad\qquad\qquad\qquad\qquad\qquad{}
   -\delta_{\alpha\delta}\Hat{\mathcal{D}}_\varepsilon\Hat{\mathcal{D}}_\varphi
   \Hat{\mathcal{D}}_\varepsilon\Hat{\mathcal{D}}_\varphi
   +\delta_{\alpha\delta}\Hat{\mathcal{D}}_\varepsilon\Hat{\mathcal{D}}^2
   \Hat{\mathcal{D}}_\varepsilon
   \Bigr]\biggr\}+O(t^{1+\epsilon})
\notag\\
   &=\frac{g_0^2}{(4\pi)^2}\frac{(8\pi t)^\epsilon}{1-\epsilon/2}
   \frac{1}{16t^2}\delta_{\alpha\delta}\delta_{\beta\gamma}\dim G
\notag\\
   &\qquad{}
   +\frac{g_0^2}{(4\pi)^2}\frac{(8\pi t)^\epsilon}{\epsilon}
   \tr\left[
   -\delta_{\alpha\delta}\Hat{\mathcal{F}}(x)^2_{\beta\gamma}
   -\Hat{\mathcal{F}}(x)_{\alpha\delta}
   \Hat{\mathcal{F}}(x)_{\beta\gamma}
   -\frac{1}{6}\delta_{\beta\gamma}\Hat{\mathcal{F}}(x)^2_{\alpha\delta}
   +\frac{1}{24}\delta_{\alpha\delta}\delta_{\beta\gamma}
   \Hat{\mathcal{F}}(x)^2_{\rho\rho}
   \right]
\notag\\
   &\qquad\qquad{}+O(t^{1+\epsilon}),
\label{eq:(2.23)}
\end{align}
where we have repeatedly used the fundamental relation
$[\Hat{\mathcal{D}}_\mu,\Hat{\mathcal{D}}_\nu]=\Hat{\mathcal{F}}_{\mu\nu}$.
Finally, taking the contraction
with~$\mathcal{P}_{\mu\alpha,\nu\delta,\beta\gamma}$,
\begin{align}
   &\frac{g_0^2}{(4\pi)^2}\frac{1-2\epsilon/3}{1-\epsilon/2}(8\pi t)^\epsilon
   \frac{3}{8t^2}\dim G\delta_{\mu\nu}
\notag\\
   &\qquad{}
   +\frac{g_0^2}{(4\pi)^2}\frac{(8\pi t)^\epsilon}{\epsilon}
   \tr\left[
   \left(-\frac{1}{3}+\frac{7}{3}\epsilon\right)
   \Hat{\mathcal{F}}(x)^2_{\mu\nu}
   +\left(-\frac{11}{12}-\frac{1}{6}\epsilon\right)
   \delta_{\mu\nu}\Hat{\mathcal{F}}(x)^2_{\rho\rho}\right]+O(t^{1+\epsilon}).
\label{eq:(2.24)}
\end{align}

Equation~\eqref{eq:(2.17)} is then given by combining Eq.~\eqref{eq:(2.24)} and
twice~Eq.~\eqref{eq:(2.20)}:
\begin{align}
   &\frac{g_0^2}{(4\pi)^2}\frac{1-2\epsilon/3}{1-\epsilon/2}(8\pi t)^\epsilon
   \frac{3}{8t^2}\dim G\delta_{\mu\nu}
\notag\\
   &\qquad{}
   +\frac{g_0^2}{(4\pi)^2}\frac{(8\pi t)^\epsilon}{\epsilon}
   \tr\left[
   \left(\frac{11}{3}+\frac{7}{3}\epsilon\right)
   \Hat{\mathcal{F}}(x)^2_{\mu\nu}
   +\left(-\frac{11}{12}-\frac{1}{6}\epsilon\right)
   \delta_{\mu\nu}\Hat{\mathcal{F}}(x)^2_{\rho\rho}\right]+O(t^{1+\epsilon}).
\label{eq:(2.25)}
\end{align}
Finally, we note $\tr[\Hat{\mathcal{F}}(x)^2_{\mu\nu}]%
=C_2(G)\Hat{F}_{\mu\rho}^a(x)\Hat{F}_{\nu\rho}^a(x)$.
% Finally, we note
% \begin{equation}
%    \tr\left[\Hat{\mathcal{F}}(x)_{\mu\nu}^2\right]
%    =\Hat{\mathcal{F}}_{\mu\rho}^{ab}(x)\Hat{\mathcal{F}}_{\rho\nu}^{ba}(x)
%    =f^{acb}f^{bda}
%    \Hat{F}_{\mu\rho}^c(x)\Hat{F}_{\rho\nu}^d(x)
%    =C_2(G)\Hat{F}_{\mu\rho}^a(x)\Hat{F}_{\nu\rho}^a(x),
% \label{eq:(3.26)}
% \end{equation}
Since we can use the tree-level relations~\eqref{eq:(2.9)} in the above
one-loop results, we can read off the coefficients in~Eq.~\eqref{eq:(2.8)}
for~$\epsilon\to0$ from~Eq.~\eqref{eq:(2.25)}:
\begin{equation}
     \zeta_{11}(t)
     =1+\frac{g_0^2}{(4\pi)^2}C_2(G)
     \left[\frac{11}{3}\epsilon(t)^{-1}+\frac{7}{3}\right],\qquad
     \zeta_{12}(t)
     =\frac{g_0^2}{(4\pi)^2}C_2(G)
     \left[-\frac{11}{12}\epsilon(t)^{-1}-\frac{1}{6}\right],
\label{eq:(2.26)}
\end{equation}
where $\epsilon(t)^{-1}\equiv1/\epsilon+\ln(8\pi t)$. These coefficients are
fundamental for the construction of a lattice energy--momentum tensor
in~Refs.~\cite{Suzuki:2013gza,Makino:2014taa}. The present simple calculational
scheme revealed that there were errors in the original diagrammatic calculation
in~Ref.~\cite{Suzuki:2013gza} (the diagrams in which the mistakes were made
have been identified; see~Ref.~\cite{Suzuki:2013gzaa}). See also the errata
for~Refs.~\cite{Suzuki:2013gza,Makino:2014taa,Asakawa:2013laa}.

\section{Conclusion}
\label{sec:3}
In the present work, we introduced a background-gauge-covariant gauge fixing
in the gradient flow equation. At least in the one-loop order, this formulation
allows a very efficient calculational scheme for the small flow time expansion
as we illustrated for the composite operator that is relevant to the
construction of a lattice energy--momentum tensor. Because of its efficiency,
further applications, including two-loop computation of the small flow time
expansion, are expected.

%\section*{Acknowledgments}
I would like to thank Kazuo Fujikawa, Kenji Hieda, and Hiroki Makino for
enjoyable discussions. The work of H.~S. is supported in part
by Grant-in-Aid for Scientific Research~23540330.

\end{document}